\documentclass[a4paper,11pt]{article}
\usepackage{jheppub}
\usepackage{eurosym}
\usepackage{amssymb}
\usepackage{amsfonts,cite}
\usepackage{amsmath}
\usepackage{graphicx}
\usepackage{multirow}
\usepackage{epsfig}
\usepackage[utf8]{inputenc}
\usepackage{multicol}
\usepackage{graphicx}
\usepackage{fancyhdr}
\usepackage{wrapfig}
\usepackage{placeins}
\usepackage{slashed}
\usepackage{lipsum}
\usepackage{array}
\usepackage{amscd}
\usepackage{tikz-cd}
\usepackage[thinlines]{easytable}
\usepackage{mathrsfs}
\usepackage[export]{adjustbox}
\usepackage{blindtext}

\usepackage{graphicx}
\usepackage{xcolor}
\usepackage{bm}
\usepackage{mathtools}

\newcommand{\dd}{\mathrm{d}}

\newbox\mybox

\newcommand\fverb{\setbox\mybox=\hbox\bgroup\verb}
\newcommand\fverbdo{\egroup\medskip\noindent\fbox{\unhbox\mybox}\ }
\newcommand\fverbit{\egroup\item[\fbox{\unhbox\mybox}]}

\abstract{We study a time-dependent non-Hermitian extension of the Sch\"utte-Da Provi- d\^encia spin-boson Hamiltonian with complex couplings. A time-dependent Dyson map relates the model to a Hermitian counterpart and induces a positive physical metric. Separating the positive and unitary parts of the map, we prove a no-go result for a	natural Gaussian number-squeeze-number class: for a nonvanishing linear spin-boson interaction, Hermiticity and bounded invertibility force the entire bosonic part of the Dyson map to be unitary. The	squeezing parameter therefore selects a time-dependent Hermitian frame rather than contributing to the metric.
	
The squeezed and non-squeezed Hamiltonians are related exactly by a	time-dependent unitary transformation. The conserved quantity formed from the boson number and spin projection is replaced by a transported dynamical invariant, so a closed squeezing-frame protocol cannot generate transitions between distinct invariant sectors. We then introduce an independently driven single-mode cavity with quadratic term $i\chi(t)(b^{\dagger2}-b^2)/2$. This physical drive breaks the corresponding continuous symmetry while preserving parity and couples dressed sectors differing by two bosonic quanta. The non-Hermitian asymmetry parameter, which also determines the bounded metric, tunes the effective Hermitian coupling, transition strengths and resonance 	conditions. Direct numerical propagation confirms the distinction between passive frame-induced mixing and genuine cavity transitions, the asymmetry-controlled resonance shift, and the validity of the	first-order transition formula in the weak-driving regime.}

\title{Bounded Dyson maps and cavity-driven transitions in a time-dependent non-Hermitian spin-boson model}

\author[a]{Andreas Fring}
\author[b]{Marta Reboiro}

\affiliation[a]{Department of Mathematics, City St George's, University of London,  Northampton Square, \\ London EC1V 0HB, UK}
\affiliation[b]{Institute of Physics of La Plata (IFLP), Boulevard 113 \& 63,  La Plata C.P. 1900, Argentina}

\emailAdd{a.fring@city.ac.uk}	
\emailAdd{reboiro@fisica.unlp.edu.ar}


\begin{document}
	\maketitle
	
	\pagestyle{fancy}
	\fancyhead{} 
	\fancyhead[LE,RO]{\small\itshape  Bounded Dyson maps and cavity-driven transitions in a non-Hermitian spin-boson model} 
	
	\renewcommand{\headrulewidth}{0.4pt}

\section{Introduction}

Non-Hermitian Hamiltonians have become an important tool for effective
descriptions of open quantum and wave systems, including systems with gain and
loss, dissipative couplings, {\cal PT} symmetry, exceptional points, and
non-reciprocal transport \cite{el2018non,ashida2020non,bergholtz2021ex,zhang2022rev}. In general such Hamiltonians need not possess real
spectra or unitary time evolution. However, if a non-Hermitian Hamiltonian is
related to a Hermitian Hamiltonian by a suitable Dyson map, the system may be
interpreted as a consistent quasi-Hermitian/{\cal PT}-symmetric quantum theory \cite{Bender:1998ke,Benderrev,AliI,Alirev}. In this setting the
inner product is modified by a metric operator and the physical observables are
not necessarily the operators that appear most naturally in the auxiliary
Hilbert space \cite{Urubu}.

For time-independent systems the central requirement is usually the
quasi-Hermiticity relation between a non-Hermitian Hamiltonian, a Hermitian
counterpart, and a positive metric. For time-dependent systems the situation is
more subtle. The Dyson map itself contributes to the generator of time evolution
through an additional gauge-like term. As a consequence, the non-Hermitian
Hamiltonian that generates the Schr\"odinger equation is not, in general, the
physical energy observable \cite{fringmoussa,fring2023introPTt}. Instead, the observable energy is obtained by
transforming the Hermitian Hamiltonian back with the inverse Dyson map. This
distinction is essential in time-dependent non-Hermitian quantum mechanics and
will play a central role below.

In this work we consider a time-dependent non-Hermitian extension of the
Sch\"utte-Da~Provid\^encia spin-boson model
\cite{schutte1977solvable}, see section 2.1 for its precise algebraic definition. We note that the spin-$1/2$ representation of our model (\ref{SchdP}) is
closely related to, but not identical with, the Jaynes-Cummings model
\cite{jaynes2005com}. Specifically, here the model couples a bosonic mode to an $SU(2)$
degree of freedom through the anti-Jaynes-Cummings-type interaction
$S_+b^\dagger+S_-b$. Related boson-mapping
techniques have also been employed in finite-temperature many-body contexts,
for instance within thermo-field dynamics \cite{civitarese1999boson}. Time-independent pseudo-Hermitian versions have been
studied previously, for example in \cite{reboiro2022qu}. Our first aim is
to analyse the time-dependent Dyson map and its boundedness on the full bosonic
Fock space.

The model belongs to the broader family of spin-boson and
light-matter Hamiltonians. Closely related collective models include
the Tavis-Cummings model \cite{tavis1968exact}, which is the
many-atom or collective-spin rotating-wave generalisation of the
Jaynes-Cummings model, and the Dicke model \cite{dicke1954coherence},
which also contains counter-rotating couplings. Various non-Hermitian
extensions of these models have been considered
\cite{ghosh2005exactly,baga2016exc,zhang2020exp,liu2022macro,GhoshMand}.
Non-Hermiticity is commonly introduced through imaginary couplings,
phenomenological decay rates, gain-loss terms or effective
time-independent Hamiltonians, with emphasis on real spectra,
exceptional points and non-Hermitian phase transitions.

The present construction differs in two respects. First, the
non-Hermitian Hamiltonian is related to a Hermitian counterpart by an
explicitly time-dependent, bounded Dyson map, so that the metric and
the physical energy observable must be treated dynamically. Second,
the squeezing factor occurring in the Dyson map is shown to represent
a unitary Hermitian-frame transformation, whereas genuine transitions
are generated by an independently prescribed physical parametric
cavity drive.

A related correspondence between time-dependent Dyson maps and unitary
dilation-type transformations was analysed in \cite{fermiFT}. The
central question here is whether the real squeezing factor, being
unitary, merely selects a time-dependent Hermitian frame or can
nevertheless affect the positive metric. We show that this is not
merely a consequence of one convenient ordering of the factors in the
Dyson map. For a natural enlarged number-squeeze-number Ansatz,
Hermiticity and bounded invertibility force the entire bosonic part of the
Dyson map to be unitary. We then establish the exact relation 
between the time-evolution operators and identify the transported
conserved operator associated with $Q=N-S_0$.

This result also clarifies which quadratic bosonic terms can generate genuine
transitions. A term produced solely by the time derivative of a unitary frame
does not constitute an independent physical drive. We therefore distinguish it
from a parametrically driven cavity term,
$i\chi(t)(b^{\dagger2}-b^2)/2$, whose amplitude $\chi(t)$ is part of the
physical Hamiltonian. Transitions induced by time-dependent confinement are well known in
Hermitian quantum systems. A particle in an infinite well with a moving
wall, for example, exhibits mixing between the instantaneous modes,
while time-dependent cavity boundaries or mode frequencies can
parametrically excite the quantum field, giving rise to the dynamical
Casimir effect
\cite{doescher1969infinite,law1994effective,dodonov2010current}.
Transition phenomena generated by explicitly time-dependent
non-Hermitian perturbations have also been investigated, including
asymmetric, effectively unidirectional and transitionless dynamics
\cite{longhi2017non}.

The mechanism considered here is distinct from both settings. The
quadratic term arising from a time-dependent unitary factor in the
Dyson map represents a change of Hermitian frame and, for a closed
frame protocol, cannot produce a final transition between invariant
sectors. Genuine transitions arise only after introducing the
independently prescribed physical cavity drive. The non-Hermitian asymmetry parameter
$\delta(t)$, which also determines the bounded physical metric, then
tunes the transition matrix elements and resonance conditions through
its effect on the Hermitian coupling
$g(t)=\alpha(t)e^{\delta(t)}$, but does not itself open the
intersector transition channel.

Our direct numerical propagation confirms the analytical distinction
between passive frame-induced mixing and genuine cavity-induced
transitions. A closed squeezing-frame protocol produces only transient
projections onto fixed-$Q$ sectors and no final intersector
transition, whereas an active cavity pulse leaves a nonzero final
population. Frequency scans show that the metric-generating asymmetry
parameter shifts the dressed resonance and modifies its strength, and
comparison with direct propagation establishes the range of validity
of the first-order transition formula.

The paper is organised as follows. In section 2 we introduce the time-dependent non-Hermitian extension of the spin-boson Hamiltonian with complex couplings and show how it is mapped to a Hermitian counterpart by means of a time-dependent Dyson map. In section 3 we derive
the no-go result for bounded Gaussian number-squeeze-number Dyson maps.
Section 4 establishes the exact relation between the
Hermitian frames and constructs the transported invariant.
Section 5 introduces the independently driven cavity model and its
bounded metric. In section 6 we derive the dressed spectrum and
transition amplitudes, contrast passive frame-induced mixing with
active cavity transitions, and analyse numerically the resonance
structure, its control through the non-Hermitian asymmetry parameter,
and the validity of the perturbative regime. Our conclusions are stated in section 7.

\section{Time-dependent quasi-Hermiticity and bounded Dyson maps}

\subsection{Non-Hernitian spin-boson Hamiltonian}

We generalise the Sch\"utte-Da~Provid\^encia Hamiltonian \cite{schutte1977solvable} in two ways, by allowing its coupling constants to be complex, so that it becomes in general non-Hermitian, and by introducing an explicit time-dependence
\begin{equation}
	H(t)=  \omega_f(t) S_0 +  \omega_b(t) N+ \alpha(t) S_+  b^\dagger + \beta(t) S_- b,
	\qquad  \omega_f(t),  \omega_b(t), \alpha(t), \beta(t) \in \mathbb{C} . \label{SchdP}
\end{equation}
As in the original model, $N$ is the bosonic number operator, $N:=  b^\dagger b$, with $[b, b^\dagger] =1 $ and $S_0,S_\pm$ are the generators of the $SU(2)$ algebra satisfying $[S_0, S_{\pm}] = \pm S_{\pm}$, $[S_{+}, S_{-}] = 2S_0$, $S_0^{\dagger} = S_0$ and $S_{+}^{\dagger} = S_{-}$.

Following the usual procedure for time-dependent non-Hermitian Hamiltonian systems  \cite{fringmoussa}, we attempt to solve the  time-dependent Dyson equation
\begin{equation}
	h(t)= \eta(t) H(t) \eta^{-1}(t) + i \dot{\eta}(t) \eta^{-1}(t),   \label{tDyson}
\end{equation}
for the Dyson map $\eta(t)$ with $h(t)$ required to be Hermitian. We let $|\psi(t)\rangle$ denote a state evolving according to the non-Hermitian Hamiltonian $H(t)$ and define its Hermitian counterpart by $|\phi(t)\rangle=\eta(t)|\psi(t)\rangle$. The standard inner product in the Hermitian representation then induces	the time-dependent physical inner product
	\begin{equation}
		\langle\psi(t)|\widetilde{\psi}(t)\rangle_{\rho}
	:=	\langle\psi(t)|\rho(t)|\widetilde{\psi}(t)\rangle,	\qquad	\rho(t):=\eta^\dagger(t)\eta(t),  \label{demetric}
	\end{equation}
	in the non-Hermitian representation. If $\eta(t)$ is invertible, then the metric operator $\rho(t)$ is
	positive definite, and the two representations yield identical physical
	amplitudes
	\begin{equation}
		\langle\phi(t)|\widetilde{\phi}(t)\rangle	=	\langle\psi(t)|\rho(t)|\widetilde{\psi}(t)\rangle .
	\end{equation}
	The conservation of this inner product is equivalent to the
	time-dependent quasi-Hermiticity relation
	\begin{equation}
		H^\dagger(t)\rho(t)-\rho(t)H(t)	=	i\dot{\rho}(t).
	\end{equation}
	Thus the physical metric is determined by the positive factor in the
	polar decomposition of the Dyson map. 
	
	We notice that multiplication of $\eta(t)$ from
	the left by a time-dependent unitary operator changes the Hermitian frame
	without changing $\rho(t)$. This should be distinguished from inserting
	a unitary factor between noncommuting non-unitary factors, which may
	alter the positive factor of the full Dyson map and thereby change the
	metric. 
	
Attempting to solve (\ref{tDyson}),	we choose a Gaussian Dyson-map Ansatz generated by the operators	naturally associated with the model
\begin{equation}
	\eta(t) :=  S_\kappa(t) e^{\gamma(t) N } e^{\delta(t) S_0 } ,  \quad \text{with} \,\,\, S_\kappa(t)=e^{\kappa(t)( b^{\dagger 2} -b^2)/2}, \quad \label{Dysonm}
\end{equation}
and $\kappa(t),\gamma(t), \delta(t) \in \mathbb{C} $.  The factor generated by $S_0$
allows for an asymmetric rescaling of the two spin-boson couplings,
the number-operator factor implements bosonic scaling and phase
rotation, and the quadratic generator produces squeezing. This class
is sufficiently broad to test whether squeezing can contribute to a
bounded metric while remaining explicitly tractable. Substitution into equation (\ref{tDyson}) then yields
\begin{eqnarray}
	h(t)&=& \left(   \omega_f + i \dot{\delta} \right)S_0 
	+ \left(   \omega_b +  i \dot{\gamma}    \right)
	\left[  \cosh(2 \kappa) N  - \frac{1}{2}\sinh(2\kappa) ( b^{\dagger 2} +b^2)  + \sinh^2(\kappa)  \right]  \label{hermh}   \\
	&& \!\!\! + \alpha e^{\gamma + \delta}S_+\left(  b^\dagger  \cosh \kappa  - b \sinh \kappa \right)  
	+ \beta  e^{-(\gamma + \delta )} S_- \left(  b  \cosh \kappa  - b^\dagger \sinh \kappa \right) + i  \frac{\dot{\kappa}}{2} ( b^{\dagger 2} - b^2) .  \notag
\end{eqnarray}
For notational simplicity, the common time argument of all coefficient
functions has been suppressed here and in the following intermediate formulas.
The Hamiltonian $h(t)$ is Hermitian provided the conditions
\begin{equation}
	A_f(t):=\omega_f + i \dot{\delta} \in \mathbb{R}, \qquad  
	A_b(t):=\omega_b +  i \dot{\gamma}  \in \mathbb{R}, \qquad
	\kappa  \in \mathbb{R}, \qquad 
	\beta = \alpha^*  e^{2 \Re (\gamma + \delta )} , \label{hermcond}
\end{equation}
are satisfied.

\subsection{Polar decomposition and boundedness}

In more detail, when separating the time-dependent functions into real and imaginary parts $	\gamma=\gamma_{\rm R}+i\gamma_{\rm I}$, $	\delta=\delta_{\rm R}+i\delta_{\rm I}$ and taking $\kappa$ to be real, the map \eqref{Dysonm} has the polar decomposition
\begin{equation}
	\eta_\kappa(t)=U(t)\Omega(t), \qquad U(t)=
	S_\kappa(t)e^{i\gamma_{\rm I}(t)N}	e^{i\delta_{\rm I}(t)S_0},
	\qquad  \Omega(t)=	e^{\gamma_{\rm R}(t)N}	e^{\delta_{\rm R}(t)S_0}.
	\label{polar}
\end{equation}
The operator $U(t)$ is unitary and $\Omega(t)$ is positive. Hence the metric in (\ref{demetric}) becomes
\begin{equation}
	\rho(t)=\Omega^2(t)
	=	e^{2\gamma_{\rm R}(t)N}	e^{2\delta_{\rm R}(t)S_0}.
	\label{metricgeneral}
\end{equation}
Thus the real parts of $\gamma$ and $\delta$ determine the metric, whereas
$\kappa$, $\gamma_{\rm I}$ and $\delta_{\rm I}$ specify unitary Hermitian
frames. Multiplication of a Dyson map from the left by a time-dependent unitary
operator does not change $\rho$.

On the bosonic Fock basis we have $e^{\gamma N}|n\rangle=e^{\gamma n}|n\rangle$. Therefore $e^{\gamma N}$ is bounded if and only if
$\gamma_{\rm R}\leq0$, while its inverse is bounded if and only if
$\gamma_{\rm R}\geq0$. Both are bounded only when $\gamma_{\rm R}=0$. By contrast, $e^{\delta_{\rm R}S_0}$ and its inverse are bounded for every
finite $\delta_{\rm R}$ in a finite-dimensional spin representation.

The bounded regime relevant below is therefore
\begin{equation}
	\gamma\in i\mathbb R,\qquad	\delta\in\mathbb R,\qquad	\kappa\in\mathbb R,
	\label{eq:boundedregime}
\end{equation}
with
\begin{equation}
	\rho(t)=e^{2\delta(t)S_0},	\qquad	\beta(t)=\alpha^*(t)e^{2\delta(t)}.
	\label{eq:boundedmetric}
\end{equation}
A purely imaginary $\gamma$ is only a number-dependent phase rotation. In the
remainder we work in the representative $\gamma=0$, since this unitary phase
does not alter the metric or the conclusions below.

\section{A no-go result for bounded number-squeeze-number maps}

The preceding result raises the question whether a different ordering can make
the squeezing parameter enter a bounded metric. We therefore consider the
enlarged Ansatz
\begin{equation}
	\eta_{\rm g}(t)	=	e^{\gamma(t)N}S_\kappa(t)e^{\tau(t)N}e^{\delta(t)S_0},
	\qquad	\kappa(t)\in\mathbb R,	\label{etageneral}
\end{equation}
with $\gamma,\tau,\delta\in\mathbb C$. Its metric is
\begin{equation}
	\rho_{\rm g}	=	e^{\tau^*N}	S_\kappa^\dagger e^{2\gamma_{\rm R}N}S_\kappa	e^{\tau N}e^{2\delta_{\rm R}S_0}.
	\label{rhogen}
\end{equation}
Formally, $\kappa$ enters this expression whenever
$\gamma_{\rm R}\neq0$. In the interaction-sector the Hermiticity conditions exclude
this possibility in the nontrivial linear model.

The transformed interaction terms are
\begin{align}
	\eta_{\rm g}(\alpha S_+b^\dagger)\eta_{\rm g}^{-1}
	={}&
	\alpha e^{\delta+\tau+\gamma}\cosh\kappa\,S_+b^\dagger	-	\alpha e^{\delta+\tau-\gamma}\sinh\kappa\,S_+b,
	\label{eq:transplus}\\
	\eta_{\rm g}(\beta S_-b)\eta_{\rm g}^{-1}
	={}&
	\beta e^{-\delta-\tau-\gamma}\cosh\kappa\,S_-b	-	\beta e^{-\delta-\tau+\gamma}\sinh\kappa\,S_-b^\dagger.
	\label{transminus}
\end{align}
Hermiticity of the pair $S_+b^\dagger$, $S_-b$ and the pair $S_+b$, $S_-b^\dagger$ requires
\begin{equation}
	\beta	=	\alpha^*	e^{2(\delta_{\rm R}+\tau_{\rm R}+\gamma_{\rm R})}, \qquad
	\beta	=	\alpha^*	e^{2(\delta_{\rm R}+\tau_{\rm R}-\gamma_{\rm R})},
	\label{paironetwo}
\end{equation}
respectively. The gauge term $i\dot\eta_{\rm g}\eta_{\rm g}^{-1}$ contains no terms
proportional to $S_\pm$ and therefore cannot modify these conditions.

For $\alpha\neq0$ and $\sinh\kappa\neq0$, equation (\ref{paironetwo}) implies 
$ \gamma_{\rm R}=0$. Substituting this into equation \eqref{rhogen} gives $	\rho_{\rm g}	=	e^{2\tau_{\rm R}N}e^{2\delta_{\rm R}S_0}$, so that the squeezing parameter drops out. Furthermore, boundedness of both
$e^{\tau N}$ and $e^{-\tau N}$ on the full Fock space requires $\tau_{\rm R}=0$. We summarise this as follows:

\medskip

\noindent\textbf{Proposition:}
Let the spin-boson Hamiltonian contain a nonvanishing linear interaction
$\alpha S_+b^\dagger+\beta S_-b$ with scalar couplings, and let its Dyson map
be of the form \eqref{etageneral} with real, nontrivial $\kappa$. If the
Dyson-transformed Hamiltonian is Hermitian and both $\eta_{\rm g}$ and
$\eta_{\rm g}^{-1}$ are bounded on the full bosonic Fock space, then
\begin{equation}
	\gamma_{\rm R}=\tau_{\rm R}=0.
\end{equation}
Hence the entire bosonic part of the Dyson map is unitary and the physical
metric is independent of $\kappa$.

\medskip

The result applies on every time interval on which
$\alpha(t)\neq0$ and $\sinh\kappa(t)\neq0$, and it extends by continuity
through isolated zeros of $\kappa$. It is not a no-go theorem for arbitrary
Dyson maps, but it applies to the natural Gaussian number-squeeze-number class
\eqref{etageneral}, with scalar couplings and on the full Fock space.
Within this class, however, the squeezing parameter cannot enter a bounded and
boundedly invertible metric while preserving the linear
Sch\"utte-Da~Provid\^encia interaction.

\section{Unitary squeezing frames and the transported invariant}

Next we define the Dyson map without the left squeezing factor by
\begin{equation}
	\eta_0(t)=e^{\gamma(t)N}e^{\delta(t)S_0},
\end{equation}
and let $h_0(t)$ be the corresponding Hermitian Hamiltonian. Since
$\eta_\kappa=S_\kappa\eta_0$, the two Hermitian representatives satisfy the
exact identity
\begin{equation}
	h_\kappa(t)	=	S_\kappa(t)h_0(t)S_\kappa^{-1}(t)	+i\dot S_\kappa(t)S_\kappa^{-1}(t).
	\label{frameidentity}
\end{equation}
Consequently, the corresponding time-evolution operators are related by
\begin{equation}
	{\cal U}_\kappa(t,0)	=	S_\kappa(t){\cal U}_0(t,0)S_\kappa^{-1}(0).
	\label{propagatoridentity}
\end{equation}

For $\gamma=0$ the non-squeezed Hamiltonian has the form
\begin{equation}
	h_0(t)
	=	A_f(t)S_0+A_b(t)N	+g(t)S_+b^\dagger+g^*(t)S_-b,
	\qquad	g(t)=\alpha(t)e^{\delta(t)}.
	\label{h0}
\end{equation}
It commutes at every time with the operator
\begin{equation}
	Q:=N-S_0.
	\label{eq:Q}
\end{equation}
Therefore ${\cal U}_0(t,0)$ preserves each $Q$ sector. In the squeezed frame
the corresponding conserved quantity is
\begin{equation}
	Q_\kappa(t)=S_\kappa(t)QS_\kappa^{-1}(t),
	\label{Qkappa}
\end{equation}
which satisfies the dynamical-invariant equation
\begin{equation}
	\frac{\partial Q_\kappa}{\partial t}	+i[h_\kappa(t),Q_\kappa(t)]=0.
	\label{nvariant}
\end{equation}

Thus the failure of the fixed operator $Q$ to commute with
$h_\kappa(t)$ is representation-specific. The corresponding conserved
quantity in the squeezed frame is the transported invariant
$Q_\kappa(t)$. A state belonging to one eigenspace of $Q_\kappa(t)$ may
appear as a superposition of several eigenspaces of the fixed operator
$Q$, but this fixed-basis mixing does not represent a physical
transition between distinct invariant sectors.

In particular, for a closed squeezing-frame protocol satisfying
\begin{equation}
	S_\kappa(T)=S_\kappa(t_0),
\end{equation}
the initial and final Hermitian frames coincide, and
\begin{equation}
	\mathcal U_\kappa(T,t_0)
	=	S_\kappa(t_0)	\mathcal U_0(T,t_0)	S_\kappa^{-1}(t_0).
\end{equation}
The final evolution is therefore unitarily equivalent to the
$Q$-preserving evolution generated by $h_0(t)$. In the particularly
transparent case
$S_\kappa(t_0)=S_\kappa(T)=I$, one has
\begin{equation}
	\mathcal U_\kappa(T,t_0)	=	\mathcal U_0(T,t_0).
\end{equation}
Consequently, a time-dependent squeezing factor occurring solely as
part of the Hermitian-frame transformation cannot generate physical
transitions between different invariant sectors.

\section{Parametrically driven quasi-Hermitian cavity model}

\subsection{Physical Hamiltonian and bounded metric}

To obtain a genuine quadratic drive, we now introduce an independent real
cavity amplitude $\chi(t)$ as part of the physical Hamiltonian. In the
non-Hermitian representation we take
\begin{equation}
	H_{\rm cav}(t)
	={}	\omega_f(t)S_0+\omega_b(t)N
	+\alpha(t)S_+b^\dagger+\beta(t)S_-b	+	i\frac{\chi(t)}{2}(b^{\dagger2}-b^2),	\qquad \chi(t)\in\mathbb R.
	\label{eq:Hcav}
\end{equation}
We use the bounded positive Dyson map
\begin{equation}
	\eta_0(t)=e^{\delta(t)S_0},
	\qquad \delta(t)\in\mathbb R.
	\label{eq:eta0cav}
\end{equation}
The quadratic cavity term acts only in the bosonic sector and commutes with
$\eta_0$. The Hermitian partner is
\begin{equation}
	h_{\rm cav}(t)	={}	A_f(t)S_0+A_b(t)N	+g(t)S_+b^\dagger+g^*(t)S_-b	+	i\frac{\chi(t)}{2}(b^{\dagger2}-b^2),
	\label{eq:hcav}
\end{equation}
provided
\begin{equation}
	A_f=\omega_f+i\dot\delta\in\mathbb R,	\qquad	A_b = \omega_b \in\mathbb R,
	\qquad	g=\alpha e^\delta,	\qquad	\beta=\alpha^*e^{2\delta}.
\end{equation}
The physical metric is
\begin{equation}
	\rho(t)=e^{2\delta(t)S_0},
	\label{eq:cavmetric}
\end{equation}
which is positive, bounded and boundedly invertible in every finite-dimensional
spin representation.

The operator identities are initially understood on the common dense
invariant domain consisting of finite linear combinations of spin and
bosonic number states. In the numerical and perturbative analysis below we restrict to the stable regime
\begin{equation}
	A_b(t)-|\chi(t)|	\geq\varepsilon_b>0,	\qquad t\in[0,T],   \label{stabcond}
\end{equation}
and remain away from ordinary degeneracies unless explicitly stated. The condition (\ref{stabcond}) follows from the quadratic bosonic part
\begin{equation}
	h_b(t)	=	A_b(t)N	+	i\frac{\chi(t)}{2}	\left(b^{\dagger2}-b^2\right),
\end{equation}
which, in terms of canonical variables
$b=(x+ip)/\sqrt{2}$, becomes
\begin{equation}
	h_b(t)	=	\frac12	\left[	A_b(t)x^2+A_b(t)p^2+\chi(t)\{x,p\}	\right]	-\frac{A_b(t)}2.
\end{equation}
The corresponding quadratic-form matrix has eigenvalues
$A_b(t)\pm\chi(t)$ and is positive definite precisely when
$A_b(t)>|\chi(t)|$. This stability
requirement is distinct from the channel-dependent smallness
condition needed for the perturbative transition formulas.

The roles of the parameters are now distinct, with $\delta(t)$ being a metric-compatible modulation of the effective coupling and $\chi(t)$ being the physical parametric cavity drive. When $\chi=0$, $Q$ remains conserved even if $\delta$ varies. When
$\chi\neq0$,
\begin{equation}
	[Q,h_{\rm cav}(t)]	=	i\chi(t)(b^{\dagger2}+b^2)\neq0,
	\label{Qbreak}
\end{equation}
so different $Q$ sectors are genuinely coupled. Although the cavity term breaks the continuous $Q$ symmetry, it
preserves the associated parity. For a spin-$j$ representation we
define the operator
\begin{equation}
	\Pi_j	:=	\exp\left[	i\pi\left(Q+j\right)	\right]
	=	\exp\left[	i\pi\left(N-S_0+j\right)	\right].	\label{parity}
\end{equation}
Since $N-S_0+j$ has an integer spectrum, $\Pi_j$ is a Hermitian
involution
\begin{equation}
	\Pi_j^\dagger=\Pi_j,	\qquad	\Pi_j^2=I.
\end{equation}
Its action on the elementary operators is
\begin{equation}
	\Pi_j b\Pi_j^{-1}=-b,
	\qquad
	\Pi_j b^\dagger\Pi_j^{-1}=-b^\dagger,
	\qquad
	\Pi_j S_\pm\Pi_j^{-1}=-S_\pm,
\end{equation}
while $N$ and $S_0$ remain invariant. Consequently,
\begin{equation}
	[\Pi_j,h_{\rm cav}(t)]=0.
\end{equation}
The quadratic cavity term changes the eigenvalue of $Q$ by $\pm2$ and
therefore leaves its parity invariant.

For spin $1/2$ this operator becomes
\begin{equation}
	\Pi	=	\exp\left[	i\pi\left(Q+\frac12\right)	\right]	=	(-1)^N\,2S_0.
\end{equation}
The dressed sector $\mathcal H_n$ has parity
\begin{equation}
	\Pi|\psi_n^\sigma\rangle	=	(-1)^{n+1}|\psi_n^\sigma\rangle,
\end{equation}
whereas the isolated state $|\uparrow,0\rangle$ has positive parity.
Thus the cavity Hamiltonian decomposes into the two invariant chains
\begin{equation}
	\mathcal H_0\oplus\mathcal H_2\oplus\mathcal H_4\oplus\cdots  \quad \text{and} \quad 
	\operatorname{span}\{|\uparrow,0\rangle\}	\oplus\mathcal H_1\oplus\mathcal H_3\oplus\cdots .
\end{equation}

\subsection{Cavity interpretation}

The quadratic term has a standard interpretation in the effective
Hamiltonian description of a parametrically driven cavity
\cite{law1994effective,dodonov2010current}. Accordingly, for a single cavity mode
with instantaneous frequency $\omega_c(t)$,  the quadratic term may be parameterised as
\begin{equation}
	\chi(t)=\frac{\dot\omega_c(t)}{2\omega_c(t)}.
	\label{chiomega}
\end{equation}
For an idealised one-dimensional cavity with
$\omega_c(t)\propto L^{-1}(t)$ this becomes
\begin{equation}
	\chi(t)=-\frac{\dot L(t)}{2L(t)}.
\end{equation}
In a literal moving-cavity implementation, $A_b(t)=\omega_c(t)$ and the
spin-field coupling may also inherit an $L(t)$ dependence. More generally, equation \eqref{eq:hcav} may be viewed as an effective parametrically driven
single-mode model. The single-mode approximation presupposes that one selected
mode dominates and that other cavity modes are sufficiently off resonance.

The crucial distinction from equation \eqref{frameidentity} is that
$\chi(t)$ is fixed by a physical cavity-driving protocol and is
independent of the unitary Dyson-map frame parameter $\kappa(t)$. In a
literal time-dependent-frequency cavity, $\chi(t)$ and $\omega_c(t)$
are related by equation \eqref{chiomega}, whereas in a more general
effective parametrically driven mode, $\chi(t)$ may be treated as the
externally prescribed pump amplitude.

\subsection{Physical energy observable}

As argued in \cite{fringmoussa}, the physical energy observable associated with the time-dependent Dyson map is not the non-Hermitian Hamiltonian $H(t)$ itself, but rather
\begin{equation}
	\widetilde H_{\rm cav}(t)	=	\eta_0^{-1}(t)h_{\rm cav}(t)\eta_0(t)
	=	H_{\rm cav}(t)+i\eta_0^{-1}(t)\dot\eta_0(t).
\end{equation}
Explicitly, we have
\begin{equation}
	\widetilde H_{\rm cav}(t)	={}	A_f(t)S_0+A_b(t)N
	+\alpha(t)S_+b^\dagger+\beta(t)S_-b+	i\frac{\chi(t)}{2}(b^{\dagger2}-b^2).
\end{equation}
At every fixed time, $\widetilde H_{\rm cav}$ is similar to the
Hermitian operator $h_{\rm cav}$ and therefore has a real instantaneous
spectrum in the bounded regime. Ordinary degeneracies may occur, but
not defective exceptional points. This contrasts with non-Hermitian
Jaynes-Cummings-type models with an imaginary coupling, where the
block frequency contains a difference of squares and may vanish at a
finite coupling, producing exceptional points
\cite{frith2020exotic}. Here, by contrast, $\Omega_n^2=(A_b+A_f)^2+4|g|^2(n+1)$ is a sum of
non-negative terms.

\section{Dressed spectrum and cavity-induced transitions}

\subsection{Undriven dressed sectors}

We now decompose the cavity Hamiltonian as
\begin{equation}
	h_{\rm cav}(t)=h_0(t)+V_\chi(t),
	\qquad
	V_\chi(t)=
	i\frac{\chi(t)}{2}(b^{\dagger2}-b^2),
	\label{split}
\end{equation}
where $h_0$ is given by equation \eqref{h0}. For spin $1/2$, the eigenspace of
$Q$ with eigenvalue $q_n=n+1/2$ is $
	{\cal H}_n
	=
	\operatorname{span}\{
	|\downarrow,n\rangle,
	|\uparrow,n+1\rangle
	\},n=0,1,2,\ldots$.
The restriction of $h_0$ to ${\cal H}_n$ is
\begin{equation}
	h_0^{(n)}(t)
	=
	\begin{pmatrix}
		A_b n-\frac{A_f}{2} & g\sqrt{n+1}\\
		g^*\sqrt{n+1} & A_b(n+1)+\frac{A_f}{2}
	\end{pmatrix}.
	\label{eq:block}
\end{equation}
Its instantaneous eigenvalues are
\begin{equation}
	E_n^\pm(t)
	=
	A_b(t)\left(n+\frac12\right)
	\pm\frac{\Omega_n(t)}{2} 	
	\label{eq:En}
\end{equation}
with 
\begin{equation}
	  \Omega_n(t) =\sqrt{[A_b(t)+A_f(t)]^2	+4|g(t)|^2(n+1)  }.
\end{equation}
There is also the isolated state $|\psi_{\rm vac}\rangle:=|\uparrow,0\rangle$ with energy
$A_f/2$, which does not belong to any of the two-dimensional sectors $\mathcal H_n$.

Writing $g=|g|e^{i\phi}$, the dressed states may be chosen as
\begin{equation}
	|\psi_n^+\rangle	=	c_n|\downarrow,n\rangle	+e^{-i\phi}s_n|\uparrow,n+1\rangle,
\qquad 
	|\psi_n^-\rangle	=	-s_n|\downarrow,n\rangle	+e^{-i\phi}c_n|\uparrow,n+1\rangle,
	\label{eq:dressed}
\end{equation}
where $c_n=\cos\theta_n$, $s_n=\sin\theta_n$, and
\begin{equation}
	\cos(2\theta_n)
	=
	-\frac{A_b+A_f}{\Omega_n},
	\qquad
	\sin(2\theta_n)
	=
	\frac{2|g|\sqrt{n+1}}{\Omega_n}.
	\label{eq:theta}
\end{equation}

It is convenient to define
\begin{equation}
	u_n^+=c_n,\qquad v_n^+=s_n,
	\qquad
	u_n^-=-s_n,\qquad v_n^-=c_n.
\end{equation}
The upward matrix elements of $b^{\dagger2}$ are then
\begin{equation}
	B_n^{\tau\sigma}(t)
	:=
	u_{n+2}^\tau u_n^\sigma\sqrt{(n+1)(n+2)}
	+
	v_{n+2}^\tau v_n^\sigma\sqrt{(n+2)(n+3)}.
	\label{eq:Bgeneral}
\end{equation}
For example,
\begin{equation}
	B_n^{++}
	=
	c_{n+2}c_n\sqrt{(n+1)(n+2)}
	+
	s_{n+2}s_n\sqrt{(n+2)(n+3)}.
	\label{eq:Bpp}
\end{equation}
Similarly, for $n\geq2$ we define
\begin{equation}
	C_n^{\tau\sigma}(t)
	:=
	u_{n-2}^\tau u_n^\sigma\sqrt{n(n-1)}
	+
	v_{n-2}^\tau v_n^\sigma\sqrt{n(n+1)}.
	\label{eq:Cgeneral}
\end{equation}
The nonvanishing perturbation matrix elements are
\begin{equation}
	\langle\psi_{n+2}^\tau|V_\chi|\psi_n^\sigma\rangle	=	i\frac{\chi}{2}B_n^{\tau\sigma}, \qquad 
	\langle\psi_{n-2}^\tau|V_\chi|\psi_n^\sigma\rangle	=	-i\frac{\chi}{2}C_n^{\tau\sigma}, \qquad
	\langle\psi_{\rm vac}|V_\chi|\psi_1^\sigma\rangle	=	-\frac{i\chi}{\sqrt{2}}\,	e^{-i\phi}v_1^\sigma.
	\label{eq:Mdown}
\end{equation}

\subsection{Instantaneous energy corrections}

Since $V_\chi$ changes the $Q$ eigenvalue by $\pm2$,
\begin{equation}
	\Delta E_{n,\sigma}^{(1)}
	=
	\langle\psi_n^\sigma|V_\chi|\psi_n^\sigma\rangle
	=0.
	\label{eq:firstshift}
\end{equation}
Away from ordinary degeneracies, the leading instantaneous correction is
\begin{equation}
	\Delta E_{n,\sigma}^{(2)}	={}	\frac{\chi^2}{4}	\sum_{\tau=\pm}	\frac{|B_n^{\tau\sigma}|^2}
	{E_n^\sigma-E_{n+2}^\tau}+	\Theta(n-2)\frac{\chi^2}{4}	\sum_{\tau=\pm}	\frac{|C_n^{\tau\sigma}|^2}
	{E_n^\sigma-E_{n-2}^\tau} +
	\delta_{n1}\,
	\frac{\chi^2}{2}
	\frac{|v_1^\sigma|^2}
	{E_1^\sigma-E_{\rm vac}}.
	\label{eq:secondshift}
\end{equation}
where $\Theta(n-2)$ equals one for $n\geq2$ and vanishes otherwise.
Near an ordinary degeneracy, $V_\chi$ must instead be diagonalised in the
degenerate subspace. The perturbative requirement is that the relevant
off-diagonal matrix elements remain small compared with the corresponding
instantaneous gaps.

\subsection{Numerical comparison of passive and active protocols}

Having distinguished the passive squeezing-frame contribution from an
independently prescribed physical cavity drive, we now illustrate this
difference by direct numerical propagation under the Hamiltonian $h_\kappa(t)$ in (\ref{frameidentity}). We truncate the bosonic
Fock space at $N_{\max}$ and prepare the system in the dressed state
$|\psi_0^+\rangle$ of $h_0$.

\begin{figure}[h]
	\begin{minipage}[b]{\textwidth}      
		\centering
		\includegraphics[width=0.98\textwidth]{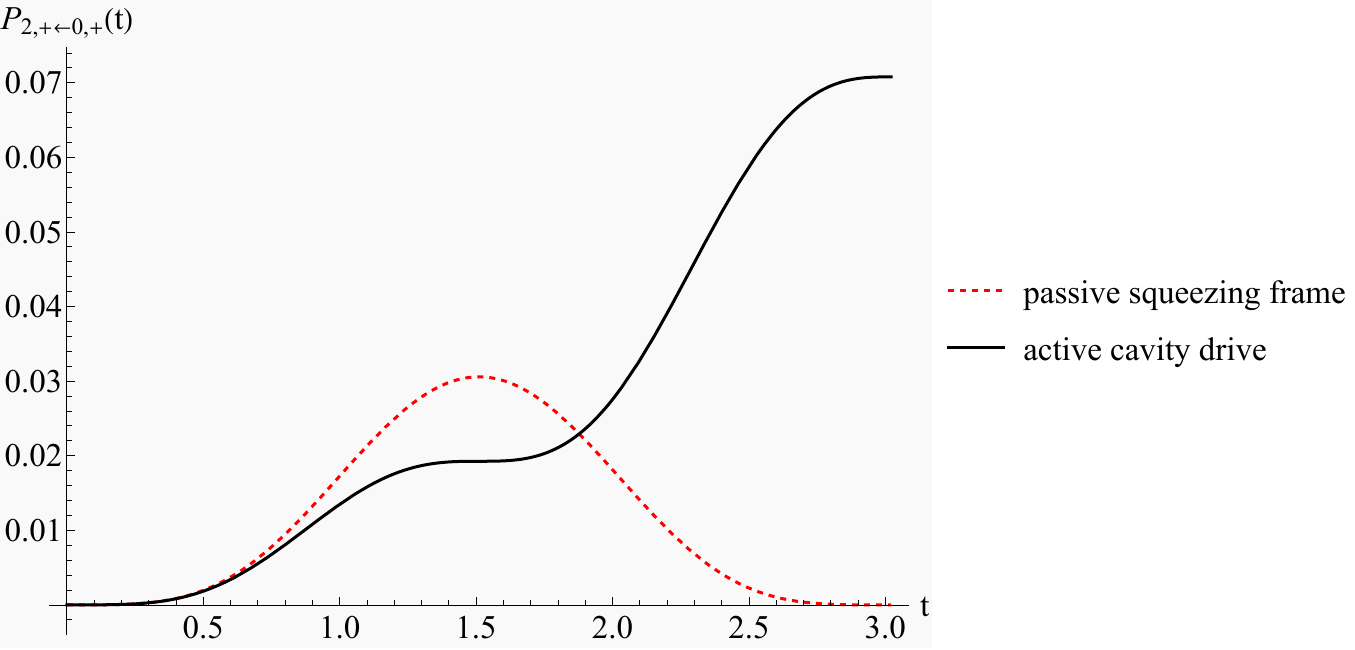}
	\end{minipage}   
	\caption{Instantaneous projection probability
		$P_{2,+\leftarrow0,+}(t)$ for the passive squeezing-frame protocol
		and the active cavity protocol. The initial state is
		$|\psi_0^+\rangle$. In the passive case the nonzero intermediate
		projection onto $|\psi_2^+\rangle$ represents fixed-frame mixing and
		returns to zero when the frame closes at $t=T$. The active cavity
		drive leaves a nonzero final population. The parameters are
		$A_b=1$, $A_f=0.4$, $g=\alpha=0.25$, $\kappa_0=0.15$,
		$E_0^+\simeq1.2433$, $E_2^+\simeq3.3231$,
		$\Delta_0^{++}\simeq2.0798$, $T\simeq3.02105$, and
		$N_{\max}=24$.}
	\label{transition02}
\end{figure}

We choose the closed protocol 
\begin{equation}
	\kappa(t)
	=
	\kappa_0
	\sin^2\left(\frac{\pi t}{T}\right),
	\qquad
	\kappa(0)=\kappa(T)=0.
\end{equation}
To distinguish this frame-induced mixing from a genuine transition, we
compare the result with the evolution under $h_{\rm cav}(t)$ in (\ref{split}) with $\chi(t) = \dot{\kappa}(t) $, that is
\begin{equation}
	h_{\rm act}(t)	=	h_0+	i\frac{\dot\kappa(t)}{2}	\left(b^{\dagger2}-b^2\right).
\end{equation}
Thus the passive and active Hamiltonians contain the same explicit
quadratic term, but only the passive Hamiltonian contains the
correlated transformation $S_\kappa h_0S_\kappa^{-1}$. Choosing
$T=2\pi/\Delta_0^{++}$ makes the active pulse resonant with the
transition $|\psi_0^+\rangle\rightarrow|\psi_2^+\rangle$. As seen in figure \ref{transition02}, the passive
protocol yields a vanishing final transition probability, whereas the
active protocol produces a nonzero population of
$|\psi_2^+\rangle$. 

For the passive protocol, the quantity shown at intermediate times is
the projection onto the fixed dressed state $|\psi_2^+\rangle$. It
should not be interpreted as a transition between eigenspaces of the
transported invariant $Q_\kappa(t)$. Since
$S_\kappa(0)=S_\kappa(T)=I$, the initial and final frames coincide and
the final value has an unambiguous transition-probability
interpretation.

All numerical calculations  are performed for spin $1/2$ by
truncating the bosonic Fock space at occupation number $N_{\max}$ and
directly integrating the time-dependent Schr\"odinger equation. Increasing the cutoff from $N_{\max}=20$ to $N_{\max}=24$ changes
the displayed probabilities by less than $10^{-11}$. Norm and parity conservation are
monitored throughout the evolution.

\subsection{Representation-independent transition probabilities}

Suppose the state is prepared at $t=0$ in
$|\psi_n^\sigma(0)\rangle$ and measured at time $T$ in the dressed basis
$|\psi_m^\tau(T)\rangle$. The Hermitian-representation probability is
\begin{equation}
	P_{m,\tau\leftarrow n,\sigma}(T)
	=
	\left|	\langle\psi_m^\tau(T)|	{\cal U}_{\rm cav}(T,0)	|\psi_n^\sigma(0)\rangle	\right|^2.
\end{equation}
The corresponding non-Hermitian states are
\begin{equation}
	|\Psi_n^\sigma(t)\rangle	=	\eta_0^{-1}(t)|\psi_n^\sigma(t)\rangle.
\end{equation}
Using $\rho=\eta_0^\dagger\eta_0$, the same probability is
\begin{equation}
	P_{m,\tau\leftarrow n,\sigma}(T)
	=	\left|	\langle\Psi_m^\tau(T)|	\rho(T)	{\cal U}_{H}(T,0)	|\Psi_n^\sigma(0)\rangle	\right|^2.
\end{equation}
Here $\mathcal U_{\rm cav}(T,0)$ denotes the time-evolution operator
generated by $h_{\rm cav}(t)$, whereas $\mathcal U_H(T,0)$ is generated
by the non-Hermitian Hamiltonian $H_{\rm cav}(t)$. Thus the transition probability is invariant under the Dyson map when states,
time-evolution operators and inner products are transformed consistently.

For a time-dependent background, the first-order amplitude should be written
using the exact block-diagonal time-evolution of $h_0(t)$:
\begin{align}
	{\cal A}^{(1)}_{m,\tau\leftarrow n,\sigma}(T)
	=	-i\int_0^T\dd t\,	\langle\psi_m^\tau(T)|	{\cal U}_{0,m}(T,t)	V_\chi(t)	{\cal U}_{0,n}(t,0)	|\psi_n^\sigma(0)\rangle.
	\label{exactblockamp}
\end{align}
Here ${\cal U}_{0,n}$ denotes the exact evolution generated by the
$2\times2$ block $h_0^{(n)}(t)$. Equation~\eqref{exactblockamp}
automatically includes all derivative couplings associated with a
time-dependent dressed basis.

For constant background parameters, the channel
$|\psi_n^\sigma\rangle\to|\psi_{n+2}^\tau\rangle$ reduces to
\begin{equation}
	{\cal A}^{(1)}_{n+2,\tau\leftarrow n,\sigma}(T)
	=	\frac{B_n^{\tau\sigma}}{2}	\int_0^T\dd t\,	\chi(t)e^{i\Delta_n^{\tau\sigma}t},
\end{equation}
where
\begin{equation}
	\Delta_n^{\tau\sigma}	=	E_{n+2}^\tau-E_n^\sigma.
\end{equation}
For constant background parameters, we first consider a monochromatic parametric pump
\begin{equation}
	\chi(t)=\chi_0\cos(\Omega t),
\end{equation}
applied over the observation interval $0\leq t\leq T$. We obtain
\begin{equation}
	{\cal A}^{(1)}_{n+2,\tau\leftarrow n,\sigma}(T)
	=	\frac{\chi_0B_n^{\tau\sigma}}{4}
	\left[	\frac{e^{i(\Delta_n^{\tau\sigma}+\Omega)T}-1}	{i(\Delta_n^{\tau\sigma}+\Omega)}
	+	\frac{e^{i(\Delta_n^{\tau\sigma}-\Omega)T}-1}	{i(\Delta_n^{\tau\sigma}-\Omega)}	\right].
\end{equation}
The transition is resonantly enhanced when
\begin{equation}
	\Omega\simeq|\Delta_n^{\tau\sigma}|.
\end{equation}

For a finite effective pump one may instead introduce a smooth
envelope, for example
\begin{equation}
	\chi_{\rm p}(t)	=	\chi_0	\sin^2\left(\frac{\pi t}{T}\right)	\cos(\Omega t), \label{envpulse}
\end{equation}
so that the active quadratic term vanishes smoothly at the initial and
final times. Such endpoint conditions do not force the transition
amplitude to vanish, since the physical drive is not generated by a
unitary change of Hermitian frame.

For a literal time-dependent-frequency cavity, the stronger closure
condition
\begin{equation}
	\omega_c(T)=\omega_c(0)	\quad\Longleftrightarrow\quad	\int_0^T\chi(t)\,dt=0
\end{equation}
must be imposed. This condition removes only the zero-frequency
component of $\chi(t)$ and does not in general eliminate its Fourier
component at the dressed-state gap. Consequently, a physically closed
cavity protocol may still produce a nonzero final transition
probability.

\subsection{Numerical validity of the first-order approximation}

\begin{figure}[h]
	\begin{minipage}[b]{\textwidth}      
		\centering
		\includegraphics[width=0.98\textwidth]{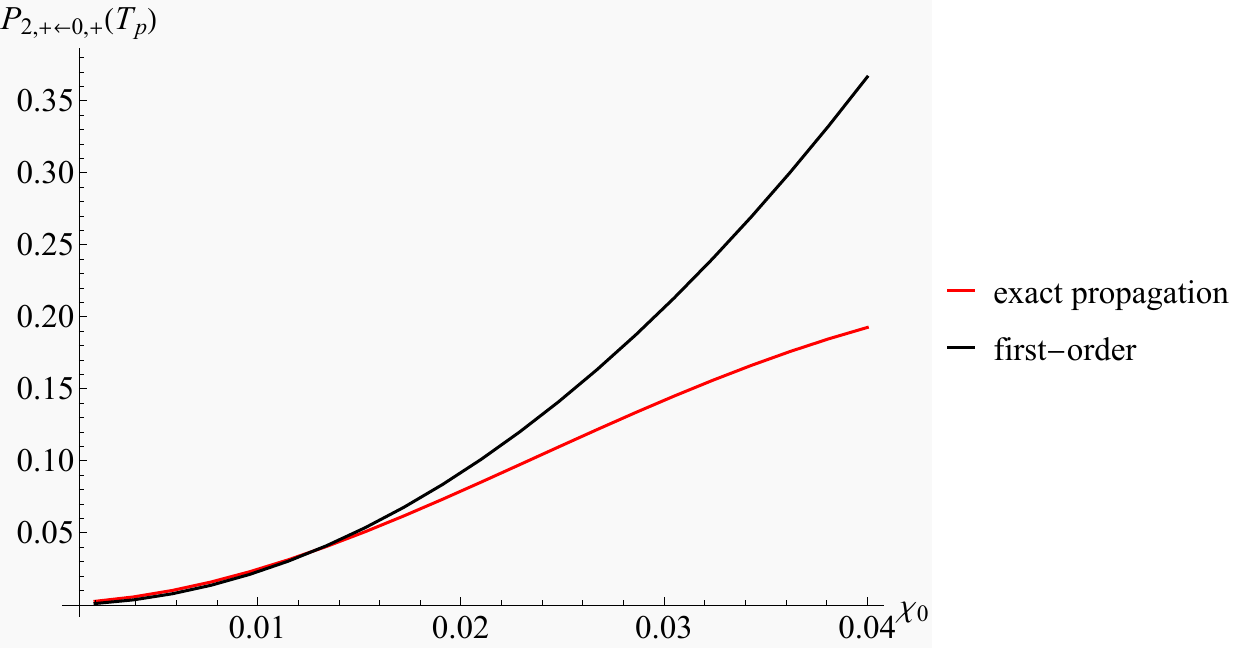}
	\end{minipage}   
	\caption{Comparison between the exact transition probability
		$P_{2,+\leftarrow0,+}(T_{\rm p})$, obtained by direct numerical
		propagation under the active cavity Hamiltonian, and the
		corresponding first-order perturbative result, shown as
		functions of the pulse amplitude $\chi_0$. The carrier
		frequency is fixed at the resonance
		$\Omega=\Delta_0^{++}$ for $\delta=0$. The parameters are $A_b=1$, $A_f=0.4$,
		$\alpha=0.25$, $T_{\rm p}=50$, and $N_{\max}=24$.}
	\label{ExactPert}
\end{figure}

Next we briefly examine the accuracy of the first-order transition formula.
We fix $\delta=0$, use the smoothly switched pulse (\ref{envpulse}) with $T=T_{\rm p}=50$ and tune the carrier frequency to the analytical resonance,
\begin{equation}
	\Omega=\Delta_0^{++}.
\end{equation}
For several values of the pulse amplitude $\chi_0$, we compare the
probability obtained by direct propagation under $h_{\rm cav}(t)$ with
the first-order prediction
\begin{equation}
	P_{2,+\leftarrow0,+}^{(1)}(T_{\rm p})
	=
	\left|	\frac{B_0^{++}}{2}	\int_0^{T_{\rm p}}\dd t\,	\chi(t)e^{i\Delta_0^{++}t}	\right|^2.
\end{equation}
Since the pulse amplitude enters linearly in $\chi(t)$, the
first-order probability is expected to scale quadratically with
$\chi_0$. The numerical results are displayed in figure~\ref{ExactPert}. 

For
small amplitudes, the exact transition probability agrees closely with
the perturbative prediction and follows the expected quadratic growth,
confirming the validity of the first-order treatment in the weak-drive
regime. As $\chi_0$ increases, the exact result gradually departs from
the perturbative curve and remains systematically smaller. This is the
expected behaviour: the first-order expression neglects depletion of
the initial dressed state as well as higher-order processes, both of
which become increasingly relevant for stronger driving. The figure therefore illustrates the weak-driving range in which the
analytical transition formula remains quantitatively accurate and its
gradual breakdown as the pulse amplitude increases. 

\subsection{Control through the metric-generating asymmetry parameter}

In the bounded quasi-Hermitian regime, the real parameter $\delta(t)$
plays two related roles. It determines the physical metric,
$\rho(t)=e^{2\delta(t)S_0}$, and, for fixed non-Hermitian coupling
$\alpha(t)$, it also determines the effective Hermitian spin-boson coupling
\begin{equation}
	|g(t)|=|\alpha(t)|e^{\delta(t)}.
	\label{gdelta}
\end{equation}
Consequently, varying $\delta(t)$ while keeping $\alpha(t)$ fixed
changes the physical Hermitian Hamiltonian through $g(t)$. It modifies
the dressed-state mixing angles, the matrix elements
$B_n^{\tau\sigma}$ and $C_n^{\tau\sigma}$, and the energy gaps
$\Delta_n^{\tau\sigma}$. The non-Hermitian asymmetry parameter can
therefore tune both the strength and the resonance condition of a
transition opened by the physical cavity drive $\chi(t)$.

This control effect should not be interpreted as a dynamical effect of
the metric alone. If $\delta(t)$ were varied while $\alpha(t)$ were
simultaneously adjusted so that $g(t)=\alpha(t)e^{\delta(t)}$ remained
fixed, the Hermitian Hamiltonian $h_{\rm cav}(t)$ would be unchanged.
The resulting change would then amount only to a different
non-Hermitian representation of the same physical dynamics, and all
transition probabilities would remain unchanged.

Moreover, $\delta(t)$ alone does not open transitions between different
$Q$ sectors. When $\chi(t)=0$, the Hamiltonian $h_0(t)$ commutes with
$Q$ for arbitrary time-dependent $\delta(t)$.
Thus the physical control mechanism has two distinct components:
$\chi(t)$ opens the intersector transition channel, whereas
$\delta(t)$ tunes the corresponding dressed matrix element and phase
accumulation through its effect on $g(t)$.

To demonstrate these two effects quantitatively, we now keep the
non-Hermitian coupling $\alpha$ fixed and perform a frequency scan for
several constant values of $\delta$. Consequently we have
\begin{equation}
	g(\delta)=\alpha e^\delta,
	\qquad
	\beta(\delta)=\alpha^*e^{2\delta}.
\end{equation}
Thus changing $\delta$ at fixed $\alpha$ changes the physical
Hermitian coupling $g(\delta)$, in addition to changing the metric
$\rho=e^{2\delta S_0}$.

For each value of $\delta$, the system is prepared in the
corresponding dressed state $|\psi_0^+(\delta)\rangle$ and driven by
the common pulse
\begin{equation}
	\chi(t)	=	\chi_0	\sin^2\left(\frac{\pi t}{T_{\rm p}}\right)	\cos(\Omega t).
	\label{pulse}
\end{equation}
At the end of the pulse we calculate
\begin{equation}
	P_{2,+\leftarrow0,+}(T_{\rm p};\delta,\Omega)
	=	\left|	\langle\psi_2^+(\delta)|	\mathcal U_{\rm cav}(T_{\rm p},0)	|\psi_0^+(\delta)\rangle	\right|^2.
	\label{probab}
\end{equation}

The leading resonance condition is
\begin{equation}
	\Omega	\simeq	\Delta_0^{++}(\delta)	=	2A_b+	\frac12
	\left[	\Omega_2(\delta)-\Omega_0(\delta)	\right],
\end{equation}
where
\begin{equation}
	\Omega_n(\delta)	=	\sqrt{	(A_b+A_f)^2+	4|\alpha|^2e^{2\delta}(n+1)	}.
\end{equation}
Varying $\delta$ at fixed $\alpha$ therefore changes both the
resonance position and the transition matrix element
$B_0^{++}(\delta)$.

\begin{figure}[h]
	\begin{minipage}[b]{\textwidth}      
		\centering
		\includegraphics[width=0.98\textwidth]{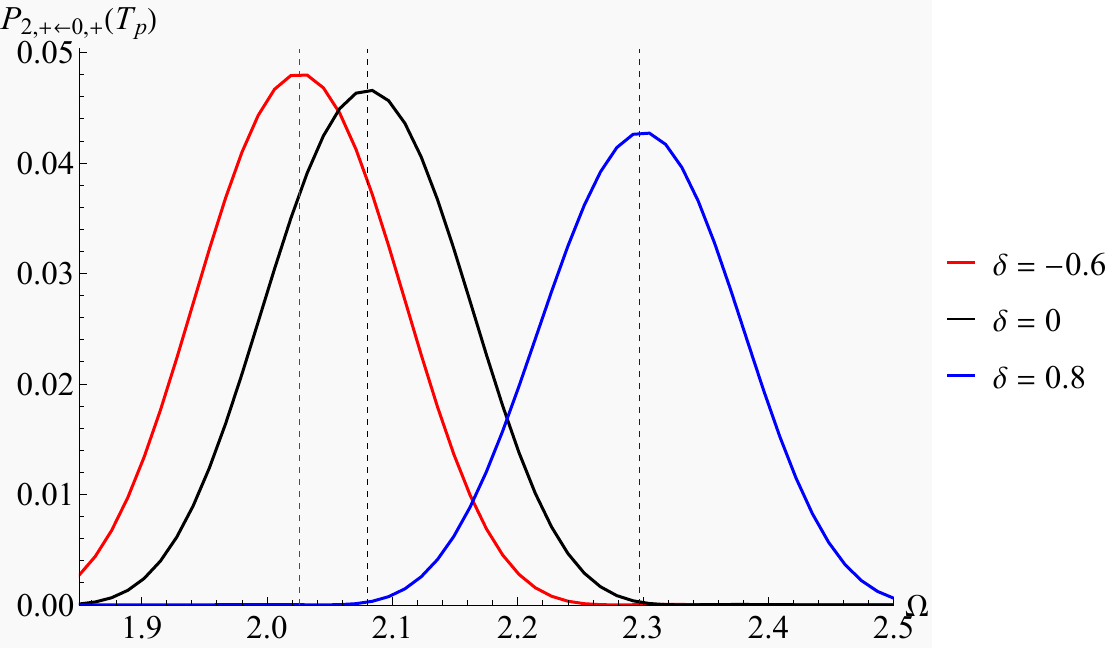}
	\end{minipage}   
	\caption{Final transition probability
		$P_{2,+\leftarrow0,+}(T_{\rm p};\delta,\Omega)$ for the active
		cavity pulse \eqref{pulse}, plotted as a function of the
		carrier frequency $\Omega$. The vertical dashed lines indicate the
		analytical resonance positions
		$\Omega=\Delta_0^{++}(\delta)$. Increasing $\delta$ at fixed
		$\alpha$ shifts the resonance to higher frequency and moderately
		reduces its peak height. The parameters are
		$A_b=1$, $A_f=0.4$, $\alpha=0.25$, $\chi_0=0.015$,
		$T_{\rm p}=50$, and $N_{\max}=24$.}
	\label{Transmetric}
\end{figure}

Figure~\ref{Transmetric} shows the final probabilities
$P_{2,+\leftarrow0,+}(T_{\rm p};\delta,\Omega)$ as functions of the
carrier frequency $\Omega$ for three fixed values of $\delta$. 

The
vertical dashed lines indicate the analytical resonance positions
$\Delta_0^{++}(\delta)$. As $\delta$ increases, the effective coupling
$g(\delta)=\alpha e^\delta$ becomes larger and the resonance moves to
higher frequencies. The numerical maxima agree with the analytical
gaps to within $0.4\%$, as summarised in
Table~\ref{gappeaks}.

\begin{table}[ht]
	\centering
	\begin{tabular}{cccccc}
		\hline	$\delta$	&		$g(\delta)$	&	$\Delta_0^{++}$	&	$\Omega_{\rm max}$	&	relative difference	&	$P_{\rm max}$
		\\	\hline
		$-0.6$	&		$0.137203$		&	$2.02592$		&		$2.032$		&		$0.30\%$		&		$0.0479562$		\\
		$0$	&	$0.25$	&		$2.07980$	&		$2.084$		&	$0.20\%$	&		$0.0465646$	\\
		$0.8$	&	$0.556385$	&	$2.29691$	&		$2.305$	&	$0.35\%$	&	$0.0427003$	\\	\hline
	\end{tabular}
	\caption{Comparison of the analytical dressed-state gaps $\Delta_0^{++}$ with the numerically determined resonance
		frequencies $\Omega_{\rm max}$. The frequency-grid spacing is $\Delta\Omega=0.013$, so the differences between the analytical gaps and the tabulated numerical maxima are
		below the numerical frequency resolution.}
	\label{gappeaks}
\end{table}

The small differences between the tabulated peak positions and the
analytical gaps are smaller than the frequency-grid spacing used in
the calculation and should therefore not be interpreted as systematic
physical shifts. The finite pulse envelope also broadens the
resonances.

The peak probability decreases moderately as $\delta$ increases. This
behaviour is consistent with
\begin{equation}
	|B_0^{++}(-0.6)|^2\simeq5.8938,	\qquad	|B_0^{++}(0)|^2\simeq5.7004,	\qquad	|B_0^{++}(0.8)|^2\simeq5.1626.
\end{equation}
The non-Hermitian asymmetry parameter $\delta(t)$, which also
determines the bounded physical metric, modifies the effective
Hermitian coupling and thereby tunes the dressed matrix elements and
resonance conditions.

\section{Conclusion}

We have constructed and analysed a time-dependent quasi-Hermitian extension of
the Sch\"utte-Da~Provid\^encia spin-boson model. The polar decomposition of
the Dyson map separates the positive metric factor from the unitary freedom in
the Hermitian representation. In the bounded regime the physical metric is
carried by the finite-dimensional spin scaling,
$\rho=e^{2\delta S_0}$, while the bosonic number rotation and real squeezing
factor are unitary.

We strengthened this conclusion by proving a no-go proposition for the natural
Gaussian number-squeeze-number class. For a nonzero linear spin-boson
interaction with scalar couplings, Hermiticity forces the number scaling to the
left of the squeeze to be unitary, while bounded invertibility forces the
remaining bosonic number scaling to be unitary as well. Thus the entire
bosonic part of a bounded map in this class is necessarily unitary, and
$\kappa$ cannot enter the physical metric.

The squeezed Hermitian Hamiltonian is consequently related exactly to the
non-squeezed one by a time-dependent unitary transformation. The fixed
operator $Q=N-S_0$ is replaced by the transported invariant
$Q_\kappa=S_\kappa QS_\kappa^{-1}$, and closed squeezing-frame protocols do
not generate physical transitions between distinct invariant sectors.

To formulate a genuine transition mechanism, we introduced an independently
parametrically driven single-mode cavity. Its quadratic term
$i\chi(b^{\dagger2}-b^2)/2$ is part of the physical Hamiltonian and genuinely
couples sectors differing by two bosonic quanta while preserving parity. The
first-order instantaneous energy correction vanishes, the leading shifts are
second order, and the first-order transition amplitudes display the expected
parametric resonance. The non-Hermitian asymmetry parameter $\delta(t)$, which also
determines the bounded physical metric, modifies the effective
Hermitian coupling and thereby tunes the dressed matrix elements and
resonance conditions, but does not itself open intersector channels.

Direct numerical propagation corroborates these conclusions. A closed passive squeezing-frame protocol produces transient projections onto fixed dressed sectors but no final intersector transition, whereas an active cavity pulse with the same explicit quadratic coefficient leaves
a nonzero final population. Frequency scans confirm that increasing
the metric-generating asymmetry parameter at fixed non-Hermitian
coupling shifts the dressed resonance and modifies its strength. The
numerical maxima agree with the analytical dressed-state gaps within
the frequency resolution used, and direct comparison with the
first-order result confirms its accuracy in the weak-driving regime
and its gradual breakdown as the pulse amplitude increases.

\newif\ifabfull\abfulltrue


\begin{thebibliography}{10}
	
	\bibitem{el2018non}
	R.~El-Ganainy, K.~G. Makris, M.~Khajavikhan, Z.~H. Musslimani, S.~Rotter, and
	D.~N. Christodoulides,
	\newblock Non-Hermitian physics and PT symmetry,
	\newblock Nat. Phys. {\bf 14}(1), 11--19 (2018).
	
	\bibitem{ashida2020non}
	Y.~Ashida, Z.~Gong, and M.~Ueda,
	\newblock Non-Hermitian physics,
	\newblock Adv. Phys. {\bf 69}(3), 249--435 (2020).
	
	\bibitem{bergholtz2021ex}
	E.~J. Bergholtz, J.~C. Budich, and F.~K. Kunst,
	\newblock Exceptional topology of non-Hermitian systems,
	\newblock Rev. Mod. Phys. {\bf 93}(1), 015005 (2021).
	
	\bibitem{zhang2022rev}
	X.~Zhang, T.~Zhang, M.-H. Lu, and Y.-F. Chen,
	\newblock A review on non-Hermitian skin effect,
	\newblock Adv. Phys.: X {\bf 7}(1), 2109431 (2022).
	
	\bibitem{Bender:1998ke}
	C.~M. Bender and S.~Boettcher,
	\newblock Real Spectra in Non-Hermitian Hamiltonians Having PT Symmetry,
	\newblock Phys. Rev. Lett. {\bf 80}, 5243--5246 (1998).
	
	\bibitem{Benderrev}
	C.~M. Bender,
	\newblock Making sense of non-Hermitian Hamiltonians,
	\newblock Rept. Prog. Phys. {\bf 70}, 947--1018 (2007).
	
	\bibitem{AliI}
	A.~Mostafazadeh,
	\newblock Pseudo-Hermiticity versus PT symmetry: The necessary condition for
	the reality of the spectrum of a non-Hermitian Hamiltonian,
	\newblock J. Math. Phys. {\bf 43}, 202--212 (2002).
	
	\bibitem{Alirev}
	A.~Mostafazadeh,
	\newblock Pseudo-Hermitian Representation of Quantum Mechanics,
	\newblock Int. J. Geom. Meth. Mod. Phys. {\bf 7}, 1191--1306 (2010).
	
	\bibitem{Urubu}
	F.~G. Scholtz, H.~B. Geyer, and F.~Hahne,
	\newblock Quasi-Hermitian Operators in Quantum Mechanics and the Variational
	Principle,
	\newblock Ann. Phys. {\bf 213}, 74--101 (1992).
	
	\bibitem{fringmoussa}
	A.~Fring and M.~H.~Y. Moussa,
	\newblock Unitary quantum evolution for time-dependent quasi-Hermitian systems
	with nonobservable Hamiltonians,
	\newblock Phys. Rev. A {\bf 93}(4), 042114 (2016).
	
	\bibitem{fring2023introPTt}
	A.~Fring,
	\newblock An introduction to PT-symmetric quantum mechanics-time-dependent
	systems,
	\newblock J. Phys.: Conf. Ser. {\bf 2448}(1), 012002 (2023).
	
	\bibitem{schutte1977solvable}
	D.~Sch{\"u}tte and J.~Da~Provi- d\^encia,
	\newblock A solvable model of boson condensation,
	\newblock Nucl. Phys. A {\bf 282}(3), 518--532 (1977).
	
	\bibitem{jaynes2005com}
	E.~T. Jaynes and F.~W. Cummings,
	\newblock Comparison of quantum and semiclassical radiation theories with
	application to the beam maser,
	\newblock Proc.  IEEE {\bf 51}(1), 89--109 (1963).
	
	\bibitem{civitarese1999boson}
	O.~Civitarese and M.~Reboiro,
	\newblock Boson mapping at finite temperature: An application to the thermo
	field dynamics,
	\newblock Phys. Rev. C {\bf 60}, 034302 (1999).
	
	\bibitem{reboiro2022qu}
	M.~Reboiro and D.~Tielas,
	\newblock Quantum work from a pseudo-Hermitian Hamiltonian,
	\newblock Quantum Rep. {\bf 4}(4), 589--603 (2022).
	
	\bibitem{tavis1968exact}
	M.~Tavis and F.~W. Cummings,
	\newblock Exact solution for an N-molecule—radiation-field Hamiltonian,
	\newblock Phys. Rev. {\bf 170}(2), 379 (1968).
	
	\bibitem{dicke1954coherence}
	R.~H. Dicke,
	\newblock Coherence in spontaneous radiation processes,
	\newblock Phys. Rev. {\bf 93}(1), 99 (1954).
	
	\bibitem{ghosh2005exactly}
	P.~K. Ghosh,
	\newblock Exactly solvable non-Hermitian Jaynes--Cummings-type Hamiltonian
	admitting entirely real spectra from supersymmetry,
	\newblock J. Phys. A: Math. Theor. {\bf 38}(33),
	7313--7323 (2005).
	
	\bibitem{baga2016exc}
	F.~Bagarello, F.~Gargano, M.~Lattuca, R.~Passante, L.~Rizzuto, and S.~Spagnolo,
	``Exceptional Points in a non-Hermitian extension of the Jaynes-Cummings Hamiltonian,''
	in \emph{Non-Hermitian Hamiltonians in Quantum Physics}, pp.~83--95,
	Springer, 2016.
	
	\bibitem{zhang2020exp}
	G.-Q. Zhang, Z.~Chen, and J.~You,
	\newblock Experimentally accessible quantum phase transition in a non-Hermitian
	Tavis-Cummings model engineered with two drive fields,
	\newblock Phys. Rev. A {\bf 102}(3), 032202 (2020).
	
	\bibitem{liu2022macro}
	N.~Liu, S.~Huang, and J.-Q. Liang,
	\newblock Macroscopic quantum states in Dicke model of PT-symmetric
	non-Hermitian Hamiltonian and superradiant phase with imaginary atomic
	population,
	\newblock Results Phys. {\bf 40}, 105813 (2022).
	
	\bibitem{GhoshMand}
	G.~Das, A.~Ghosh, and B.~P. Mandal,
	\newblock Quantum phase transitions and entanglement entropy in a non-Hermitian Jaynes-Cummings model,
	\newblock Ann. Phys. {\bf 490},  170484 (2026).
	
	\bibitem{fermiFT}
	A.~Fring and T.~Taira,
	\newblock Non-Hermitian Quantum Fermi Accelerator,
	\newblock Phys. Rev. A {\bf 108}(1), 012222 (2023).
	
	\bibitem{doescher1969infinite}
	S.~Doescher and M.~Rice,
	\newblock Infinite square-well potential with a moving wall,
	\newblock Am. J. Phys. {\bf 37}(12), 1246--1249 (1969).
	
	\bibitem{law1994effective}
	C.~Law,
	\newblock Effective Hamiltonian for the radiation in a cavity with a moving
	mirror and a time-varying dielectric medium,
	\newblock Phys. Rev. A {\bf 49}(1), 433 (1994).
	
	\bibitem{dodonov2010current}
	V.~Dodonov,
	\newblock Current status of the dynamical Casimir effect,
	\newblock Phys. Scr. {\bf 82}(3), 038105 (2010).
	
	\bibitem{longhi2017non}
	S.~Longhi and G.~Della~Valle,
	\newblock Non-Hermitian time-dependent perturbation theory: Asymmetric
	transitions and transitionless interactions,
	\newblock Ann. Phys. {\bf 385}, 744--756 (2017).
	
	\bibitem{frith2020exotic}
	T.~Frith,
	\newblock Exotic entanglement for non-Hermitian Jaynes--Cummings Hamiltonians,
	\newblock J. Phys. A: Math. Theor. {\bf 53}(48), 485303 (2020).
	
	
	
	
\end{thebibliography}
\end{document}